\documentclass[fleqn,twoside]{article}
\usepackage{espcrc2,amsfonts}
\usepackage{graphicx}
\usepackage[figuresright]{rotating}

\newcommand{\AmS}{{\protect\the\textfont2
  A\kern-.1667em\lower.5ex\hbox{M}\kern-.125emS}}

\def\phid{\phi^\dagger}
\def\tr{\mathop{\rm tr}\nolimits}
\def\nn{\nonumber}
\def\e{\varepsilon}
\title{\vspace{-5cm}
\hfill {\normalsize FERMILAB-CONF-10-244-T}\vspace{3cm}\\
MCFM for the Tevatron and the LHC}

\author{%
John M. Campbell\address[FE]{Fermilab, P.O. Box 500, Batavia, IL 60510, USA}
and R. K. Ellis\addressmark[FE]
}
\begin{document}

\begin{abstract}
A summary is given of the current status of the next-to-leading order
(NLO) parton-level integrator MCFM. Some details are given about the
Higgs + 2-jet process and the production and decay of $t \bar{t}$,
both of which have recently been added to the code. Using
MCFM, comparisons between the Tevatron running at $\sqrt{s}=2$~TeV and
the LHC running at $\sqrt{s}=7$~TeV are made for standard model
process including the production of Higgs bosons. The case for running
the Tevatron until 16fb$^{-1}$ are accumulated by both detectors is
sketched.
\vspace{1pc}
\end{abstract}

\maketitle \thispagestyle{empty}

\section{MCFM}
MCFM is a parton-level event integrator which gives results for a
series of processes, especially those containing the bosons $W,Z$ and
$H$ and heavy quarks, $c,b$ and $t$.
Most processes are included at next-to-leading order (NLO)
and include spin correlations in the decay.
Table~\ref{MCFMprocs} gives an abbreviated summary of the processes
which are currently treated by the program. Full documentation for the program
is available at ref.~\cite{MCFM}.
We will not review these processes in
detail but rather concentrate on the new features which are present in
version 5.8 which was released in April 2010.

\begin{table}
\caption{Abbreviated summary of MCFM proceses}
\label{MCFMprocs}
\begin{tabular}{ccc}
\hline
Final state & Notes & Ref. \\
\hline
\hline
\multicolumn{3}{l}{$W/Z$ processes}  \\
\hline
$W/Z$ & &\\
$WW/ZZ/WZ$       &  & \cite{Campbell:1999ah} \\
$W b\bar{b}$  & $m_b=0$ & \cite{Ellis:1998fv}\\
$Z b\bar{b}$  & $m_b=0$ & \cite{Campbell:2000bg} \\
$W/Z + 1$~jet &   &  \\
$W/Z + 2$~jets &  & \cite{Campbell:2002tg} \\
$Wc$ & $m_c\neq 0$ &  \cite{Campbell:2005bb}\\
$Zb$ & $n_f=5$  &  \cite{Campbell:2003dd} \\
$Zb$+jet & $n_f=5$ & \cite{Campbell:2005zv} \\
\hline
\hline
\multicolumn{3}{l}{$H$ processes} \\
\hline
$H$(g.f.) &  & \\
$H$+1 jet(g.f.) & &  \\
$H$+2 jets (g.f.) & $m_t \to \infty$  &
\cite{Campbell:2006xx,Campbell:2010cz,Campbell:2010gg} \\
$WH/ZH$   &  & \\
$H$ via WBF &   & \cite{Berger:2004pca} \\
$Hb$ & $n_f=5$ &  \cite{Campbell:2002zm} \\
\hline
\hline
\multicolumn{3}{l}{$t$ processes} \\
\hline
$t$ & $s$ and $t$ channel & \cite{Campbell:2004ch} \\
$t$ & $t$ channel,$n_f=4$ & \cite{Campbell:2009ss,Campbell:2009gj}\\
$Wt$  & $n_f=5$ & \cite{Campbell:2005bb} \\
$t\bar{t}$  & with $t$ decay  & \\
\hline
\hline
\multicolumn{3}{l}{Processes not present in released version} \\
\hline
$Wb$ +jet & & \cite{Campbell:2006cu,Campbell:2008hh}\\
$WW$ +jet & & \cite{Campbell:2007ev} \\
$J/\psi$ \& $\Upsilon$ & & \cite{Campbell:2007ws} \\
$\gamma N \to J/\psi$  & & \cite{Artoisenet:2009xh}  \\
\hline
\end{tabular}
\end{table}

\section{Higgs + two jets}
\begin{figure}[t]
\begin{center}
\includegraphics[scale=0.35]{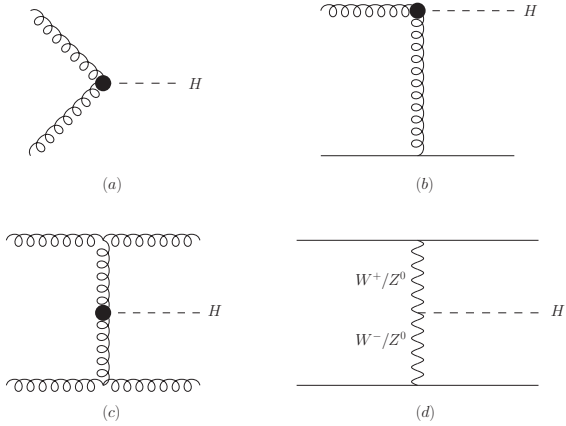}
\end{center}
\caption{Higgs production processes at lowest order.}
\label{hprod}
\end{figure}

A new process which is in MCFM version 5.8 is the production of a
Higgs boson in association with two jets. Sample diagrams contributing to
Higgs boson production are shown in Fig.~\ref{hprod}. We shall focus on
the process in Fig.~\ref{hprod}(c) and other Higgs + 4 parton processes,
which can be considered a background to the
vector boson fusion process, Fig.~\ref{hprod}(d).
The calculations underlying our implementation
are performed at NLO using an effective Lagrangian
to express the coupling of gluons to the Higgs
field~\cite{Wilczek:1977zn},
\begin{equation} \label{EffLag}
\mathcal{L}_H^{\mathrm{int}} = \frac{C}{2} \, H\,\tr
G_{\mu\nu}\,G^{\mu\nu}\, ,
\end{equation}
where the trace is over the color degrees of freedom.
At NLO the coefficient $C$
is given in the $\overline{\rm MS}$ scheme by~\cite{Djouadi:1991tka,Dawson:1990zj},
\begin{equation}
C =\frac{\alpha_S}{6 \pi v} \Big( 1 +\frac{11}{4 \pi} \alpha_S\Big)
 + {\cal O}(\alpha_S^3) \;.
\end{equation}
Here $v$ is the vacuum expectation value of the Higgs field, $v = 246$ GeV.
Phenomenological results on this process at NLO
were first published in ref.~\cite{Campbell:2006xx}
using a semi-numerical method to
calculate the virtual corrections.
Here we shall present new phenomenological results
for the Higgs + 2 jet process, based on analytic calculations of the
one-loop Higgs + 4 parton amplitudes which have recently been completed.
The use of analytic results leads to a considerable improvement
in the speed of the code.
\subsubsection{One-loop $H$ + 4 parton amplitudes}
The effective Lagrangian used in the calculation of the Higgs + 4 parton amplitudes
was simplified by introducing a complex scalar field~\cite{Dixon:2004za},
\begin{equation}
\phi = \frac{1}{2} \left( H+i A \right),\;\;\;\; \phid = \frac{1}{2} \left( H-iA \right) \;,
\end{equation}
so that the effective Lagrangian, Eq.~(\ref{EffLag}), can be written as,
\begin{eqnarray}
&&\hspace{-2em}\mathcal{L}_{H,A}^{\mathrm{int}}
= \frac{C}{2} \Big [H\,\tr G_{\mu\nu}\,G^{\mu\nu}+i A\,\tr G_{\mu\nu}\,{}^*G^{\mu\nu}\Big] \nn = \\
&=&  C \Big [\phi\,\tr G_{\scriptscriptstyle{SD}\; \mu\nu}\,G^{\mu\nu}_{\scriptscriptstyle{SD}}
            +\phid\,\tr G_{\scriptscriptstyle{ASD}\; \mu\nu}\,G^{\mu\nu}_{\scriptscriptstyle{ASD}} \Big].  \nn
\end{eqnarray}
The gluon field strength has been separated into a self-dual and an anti-self-dual component,
\begin{eqnarray}
&&\hspace{-1.5em} G_{\scriptscriptstyle{SD}}^{\mu\nu} = \frac{1}{2}
(G^{\mu\nu}+{}^*G^{\mu\nu}), G_{\scriptscriptstyle{ASD}}^{\mu\nu} = \frac{1}{2}
(G^{\mu\nu}-{}^*G^{\mu\nu})\nn \\
&&\hspace{-1.5em} {}^*G^{\mu\nu} \equiv\frac{i}{2} \e^{\mu\nu\rho\sigma} G_{\rho\sigma}\, .
\end{eqnarray}
Calculations performed in terms of the field $\phi$ are simpler than
the calculations for the Higgs boson and, moreover, the amplitudes for
$\phid$ can be obtained from the $\phi$ amplitudes by using parity.
The full Higgs boson amplitude is written as a combination of $\phi$
and $\phid$ components:
\begin{equation}
A(H)=A(\phi)+A(\phid)\; .
\end{equation}

A nice summary of all the one-loop results for the Higgs + 4 gluon
amplitudes is given in ref.~\cite{Badger:2009hw}. Full references for the
analytic calculations of the $H\bar{q}qgg$ amplitudes can be found in ref.~\cite{Badger:2009vh}.
Results for the matrix squared for the $Hq \bar{q} q \bar{q}$
process are given in ref.~\cite{Campbell:2006xx}
and for the amplitude in ref.~\cite{Dixon:2009uk}.

\subsection{Phenomenological impact}
In addition to its importance at the LHC, the Higgs + 2 jet cross section
is also important at the Tevatron. The experiments~\cite{Aaltonen:2010sv}
analyze the events with different numbers of jets separately to make
maximal use of the different kinematic structure.
In the spirit of
Ref.~\cite{Anastasiou:2009bt}, we can refine the estimate of the theoretical
uncertainty on the number of Higgs signal events originating from
QCD parton fusion processes
\begin{table}
\caption{Comparison of calculations of refs.~\cite{Anastasiou:2009bt,Campbell:2010cz}}
\label{tablecomparison}
\begin{tabular}{lccc}
ADGSW\cite{Anastasiou:2009bt}
& LO & NLO& NNLO \\
\hline
Higgs+0jet & \checkmark & \checkmark & \checkmark \\
Higgs+1jet & \checkmark & \checkmark &  \\
Higgs+2jet & \checkmark &  &  \\
\hline
\end{tabular}
\begin{tabular}{lccc}
CEW
\cite{Campbell:2010cz}
& LO & NLO& NNLO \\\hline
Higgs+0jet & \checkmark & \checkmark &  \\
Higgs+1jet & \checkmark & \checkmark &  \\
Higgs+2jet & \checkmark & \checkmark &  \\
\hline
\end{tabular}
\end{table}

In Table~\ref{tablecomparison} we contrast the two different approaches to calculating
the Higgs + 2 jet cross sections of Ref.~\cite{Anastasiou:2009bt} and Ref.~\cite{Campbell:2010cz}.
Ref.~\cite{Anastasiou:2009bt} is in essence a NNLO calculation of the total cross section, which as a
byproduct includes the Higgs + 2 jet process in leading order. The MCFM implementation~\cite{Campbell:2010cz}
only calculates the Higgs + 0 jet cross section at NLO, but also includes the Higgs + 2 jet cross section at NLO.

By using the fractions of the Higgs cross section in the
different multiplicity bins taken from Ref.~\cite{CDFnote9500}, we can
update Eq.~(4.3) of Ref.~\cite{Anastasiou:2009bt}
 (for a Higgs boson
of mass $160$~GeV) with,
\begin{eqnarray} \label{uncertain}
&&\hspace{-1.5em}\frac{\Delta N({\rm scale})}{N} = \left({^{+13.8\%}_{-15.5\%}} \right)  = \nn \\
&&\hspace{-1.5em} 60\% \cdot \left({^{ +5\%}_{ -9\%}} \right)
+29\% \cdot \left({^{+24\%}_{-23\%}} \right)
+11\% \cdot \left({^{+35\%}_{-31\%}} \right)\nn \\
\end{eqnarray}
Only the uncertainty on the Higgs~$+ \ge 2$ jet bin, (which is only 11\% of the total)
has been modified, using the results from Table~\ref{tevresults}.

\begin{table}[t]
\caption{LO and NLO Higgs + two jet cross section at $\sqrt{s}=1.96$~TeV,
together with theoretical errors.}\label{tevresults}
\begin{tabular}{llll}
\hline
$m_H$[GeV] & $\Gamma_H$[GeV] & $\sigma_{LO}$[fb] & $\sigma_{NLO}$[fb] \\
\hline
 160 & 0.0826 & $0.345^{+92\%}_{-44\%}$  & $0.476^{+35\%}_{-31\%}$ \\
\hline
\end{tabular}
\end{table}
The corresponding determination using the uncertainty  derived from the
LO result for the
Higgs~$+\ge 2$ jet bin is $(+20,0 \%, -16.9\%)$~\cite{Anastasiou:2009bt},
so the result in Eq.~(\ref{uncertain}) represents a modest improvement
in the overall theoretical error, but one which will have implications
for the Higgs search at the Tevatron.

\section{Top production and decay}
Another new process which is included at NLO is the
production of pairs of top quarks including the decay.
The top quarks are kept strictly on their mass shell, so the processes
of production and decay are separately gauge invariant, but full spin
correlations are kept.
Although this is not a new result~\cite{Bernreuther:2001rq,Melnikov:2009dn},
it is importantant to include it in the MCFM package because top pair production
is such an important background for many processes at hadron colliders.
\begin{figure}[t]
\begin{center}
\includegraphics[angle=270,scale=0.3]{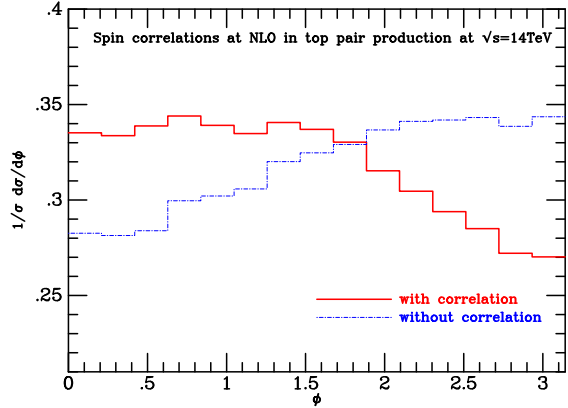}
\end{center}
\caption{Effect of spin correlations in top decay.}
\label{topphi}
\end{figure}
We can assess the importance of including these spin correlations
by looking at the angular separation of the two charged leptons coming from
top decay. The expected data sample of top quark pairs at
$\sqrt{s}=7$~TeV will be too small to observe
these correlations, but they should be observable at
$\sqrt{s} =14$~TeV.
Fig.~\ref{topphi} shows the azimuthal angle $\phi$ in the transverse plane
between the two charged leptons in top pair production events.
In addition to standard lepton and jet cuts,
$p_{T,l} > 20$~GeV, $p_{T,bjet} > 25$~GeV,
$p_{T,miss} > 40$~GeV, $\eta_l,\eta_{bjet} <2.5$
we apply the cut $p_{T,l} < 50$~GeV to constrain
the top quarks to be produced close to threshold~\cite{Mahlon:2010gw}.
These specific cuts have been suggested by Schulze~\cite{Schulze}.
 \section{Run III at the Tevatron}
\begin{figure*}[t]
\begin{center}
\includegraphics[angle=270,scale=0.5]{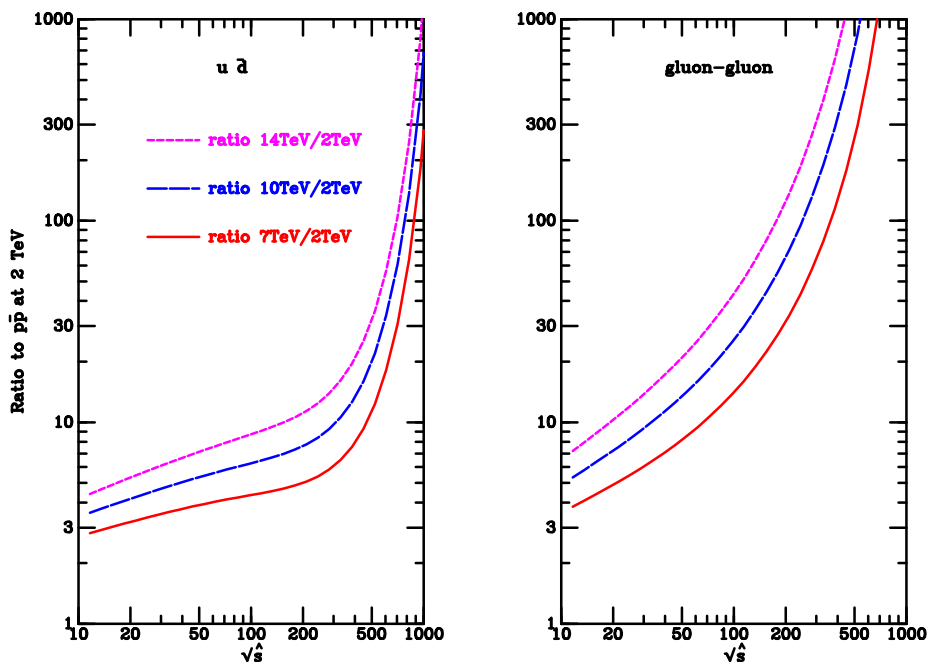}
\end{center}
\caption{Ratios of luminosities at $\sqrt{s}$=7, 10 and 14~TeV,
compared with $p \bar{p}$ at $\sqrt{s}$=2~TeV}
\label{ratcomp}
\end{figure*}

The basic ratios of cross sections are governed by the parton luminosities.
Fig.~\ref{ratcomp} shows the ratios of parton
luminosities in $pp$ collisions at $\sqrt{s}=14,10$ and $7$~TeV
compared to the  luminosity in $p\bar{p}$ collisions at $\sqrt{s}=2$~TeV.
Considering in detail the case of $7$ TeV,
in the range $\sqrt{\hat{s}} = 100-200$~GeV the $u \bar{d}$ luminosity goes up by a
factor of $4-5$ whereas the $gg$ luminosity grows by at least a factor
of $15$. This means that for $q \bar{q}$ induced processes the
Tevatron has a competitive advantage. That is, in a scenario where the
Tevatron has an accumulated luminosity of $10$~fb$^{-1}$ and the LHC
has an accumulated luminosity of $1$~fb$^{-1}$ at $\sqrt{s}=7$~TeV,
the Tevatron still has a competitive advantage for $q \bar{q} $
induced processes.

\begin{figure*}[t]
\begin{center}
\includegraphics[angle=270,scale=0.48]{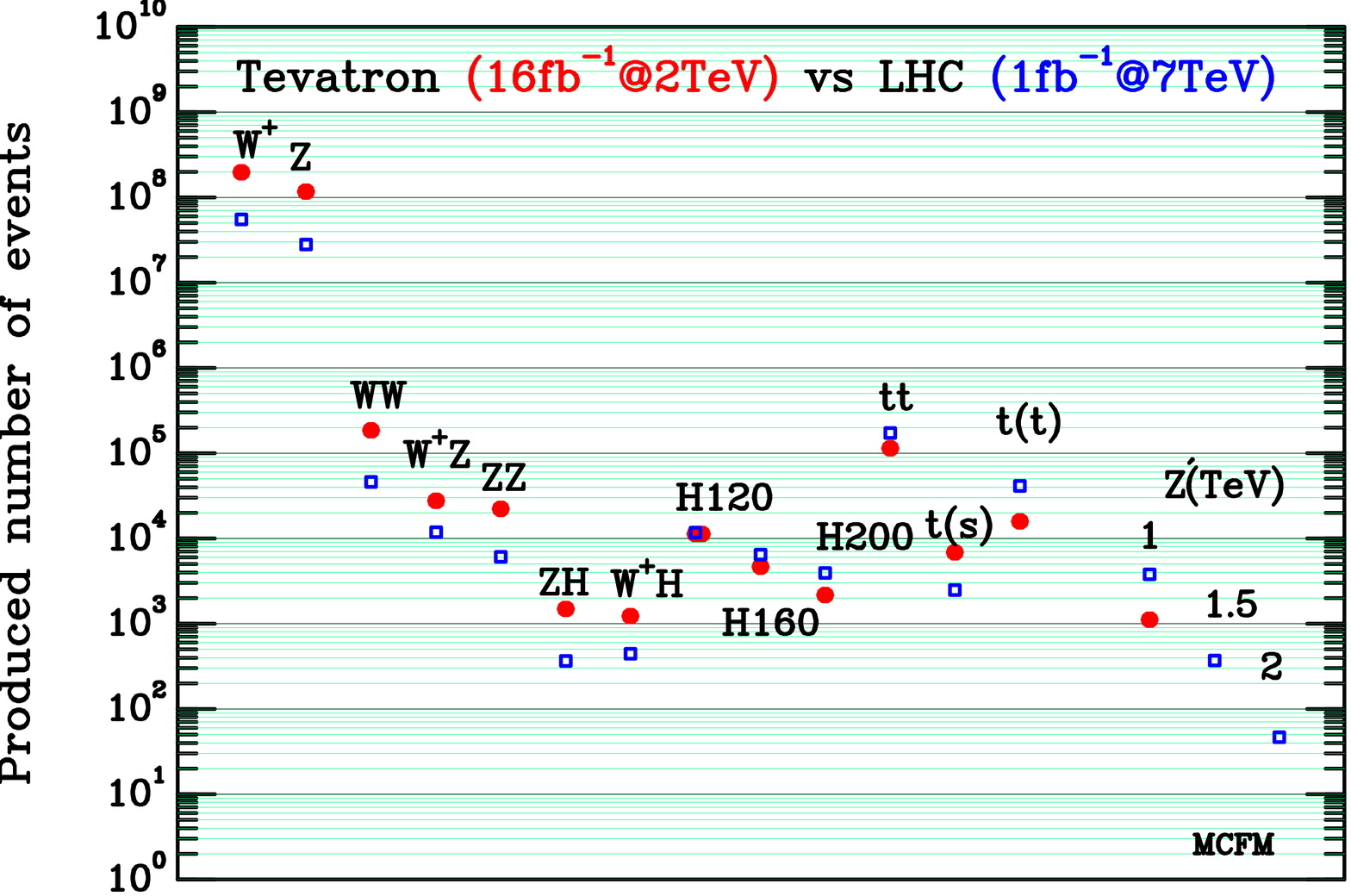}
\end{center}
\caption{Number of events for production of  $W,Z$ Higgs, top pair, single top, and $Z^\prime$ at $\sqrt{s}=2$ and $7$~TeV.
Efficiencies are assumed to be $100\%$.}
\label{rate}
\end{figure*}

We now further discuss the situation if the Tevatron were to run for three further years
after 2011; after this period the experiments would have accumulated 16~fb$^{-1}$ per experiment of analyzeable luminosity.
Fig.~\ref{rate} shows the number of events produced
for various standard model processes, assuming
16~fb$^{-1}$ of accumulated luminosity for the Tevatron
and 1~fb$^{-1}$ of accumlated luminosity for the LHC.
In this situation the Tevatron
would have a clear advantage for $q \bar{q}$ initiated processes.

The situation with regard to the standard model Higgs boson is interesting,
since at the Tevatron the low mass Higgs boson is sought in association with
a vector boson $V$ in the
$q\bar{q}$-initiated mode $q \bar{q} \to V H$.
If we take the precision standard model fits \cite{Flacher:2008zq} seriously,
the standard model 2$\sigma$-allowed region for the Higgs boson
mass is $114<M_H< 145$~ GeV. In this region
the primary decay of the Higgs boson is into $b \bar{b}$, a channel which is not
expected to be observable at the LHC until $30$~fb$^{-1}$ have been accumulated
at $\sqrt{s}=14$~TeV~\cite{Butterworth:2008iy}.

If the Tevatron were to accumulate 16~fb$^{-1}$ of analyzeable luminosity,
per experiment it could provide $3 \sigma$ evidence for the standard model Higgs boson
in the range $100 < m_H < 180$~GeV~\cite{Konigsberg}. This is an important goal,
which would provide complementary information to the information on the decay $H \to \gamma \gamma $
which will be available from $14$~TeV running at the LHC.

\section{Conclusions}
This year has been mainly a consolidation period for MCFM, but with the
introduction of two new processes at NLO, top pair production with decay and
the Higgs boson + 2 jet production at the LHC.

Using MCFM it has been shown that the Tevatron can provide important information
on $q \bar{q}$ initiated processes, and that for these processes it will be superior
to the LHC until considerable data has been accumulated at $\sqrt{s}=14$~TeV.
An important example is the Higgs boson where the $H \to b\bar{b}$ decay
of the low mass Higgs can be looked for.
Information in this channel is complementary to the information from the LHC
and probably unique until at least 2015.

\end{document}